\documentclass[a4paper,11pt]{article}
\usepackage[top=2cm,bottom=2cm,left=0.3cm,right=0.3cm]{geometry}
\usepackage{amssymb}
\usepackage{amsthm}
\usepackage{amsmath}
\usepackage{graphicx}
\usepackage[T1]{fontenc}
\usepackage{amsfonts}
\usepackage{bm}
\usepackage{enumitem}
\usepackage{subcaption}

\begin{document}
	
\title{Directed clustering in weighted networks: a new perspective}

\author{Gian Paolo Clemente$^{\dag}$,  Rosanna Grassi$^{\ddag}$ \\
	$^{\dag}$ {\small gianpaolo.clemente@unicatt.it, Department of Mathematics, Finance and Econometrics,} \\ \small{Catholic University of Milan, Italy} \\
	$^{\ddag}$ {\small rosanna.grassi@unimib.it}, \\ {\small Department of Statistics and Quantitative Methods,  University of Milano-Bicocca, Italy} \\{\small Via Bicocca degli Arcimboldi, 8, 20126 Milan, Italy Tel.+39-02-64483136}}  \maketitle


		\begin{abstract}
			In this paper, we consider the problem of assessing local clustering in complex networks. Various definitions for this measure have been proposed for the cases of networks having weighted edges, but less attention has been paid to both weighted and directed networks. 
			We provide a new local clustering coefficient for this kind of networks, starting from those existing in the literature for the weighted and undirected case.  
			Furthermore, we extract from our coefficient four specific components, in order to separately consider different link patterns of triangles.
			Empirical applications on several real networks from different frameworks and with different order are provided. The performance of our coefficient is also compared with that of existing coefficients in the literature.
		\end{abstract}
		
			
			

	\setcounter{page}{1}
	
	\newpage
	
	
	
	\section{Introduction}
	Literature in network theory mainly focused on unweighted undirected networks and several topological properties of networks have been identified through useful indicators, which enhance the efficiency of a network in carrying out its essential functionality (\cite{Barabasi}, \cite{Newman_2002}, \cite{Watts_1998}). Among these is the case of clustering coefficient that measures the tendency to which nodes in a graph tend to cluster together. Indeed, in most real networks empirical evidence shows that nodes tend to form tightly-knit groups characterized by a relatively high density of ties. In other words, the clustering coefficient is a measure of cohesion and it was developed with the aim to quantify the level to which a network manifests this property.
	
	Different definitions of clustering coefficient have been proposed for binary undirected networks (BUN). A global coefficient, often referred as transitivity, gives an overall indication of the clustering in the network being measured as the fraction of triplets (i.e. three nodes with at least two ties among them) that are closed (i.e. they form a triangle) (see \cite{Newman} and \cite{WasFaust}). A local version has been also introduced in \cite{Watts_1998} in order to quantify how close the node's neighbours are from being a clique. Although it suffers from a number of limitations\footnote{For instance, the local clustering coefficient can be biased by correlation with node's degree.}(see \cite{RavBar} and \cite{RavBar2}), the local coefficient is capable to capture the degree of social embeddedness of single nodes and for instance it is used by several mainstream indicators to assess small-world property of a network (see \cite{Watts_1998},  \cite{Walsh}, \cite{Telesford}). Unlike the local clustering coefficient, the transitivity does not suffer from the same type of limitations because it is not an average of individual fractions calculated for each node. 
	However, in many context this additional information is needed for each node. Indeed networks could be highly clustered at local level, despite showing a transitivity coefficient significantly low (see \cite{Estrada2012} pag 83). Hence a more node-oriented analysis is often required to better investigate the network cliquishness. 
	
	However, while binary networks allowed to properly model many real-world phenomena, further complexity is often needed to adequately catch heterogeneous strengths and asymmetric connections between pairs of nodes. In these contexts weighted and directed networks are fruitful tools. Furthermore, it is well known that many real-world complex systems involve non-mutual relationships, which imply non-symmetric adjacency or weighted matrices. As regard to this issue, the transitivity coefficient has been extended to both binary and weighted directed networks in \cite{Opsahl}. The proposed generalization retains the information encoded in the weights of ties.
	At the same time, local clustering coefficient have been also generalized
	to weighted undirected networks (WUN) by considering different ways to weight the neighbours of a node (see \cite{Barrat_2004}, \cite{Onnela_2005}, etc.). See \cite{Anton} and \cite{Saram} for a review of such definitions in the literature. \\
	In this context Fagiolo (\cite{Fagiolo_2007}) attempts to bridge different approaches (proposed in \cite{Milo} and \cite{Onnela_2005}) in order to present a unifying framework for computing local clustering for weighted directed networks (WDN). In addition to the measures already discussed in \cite{Milo} and \cite{Onnela_2005}, the coefficient proposed in \cite{Fagiolo_2007} allows to explicitly account for directed and weighted links and to define a specific clustering coefficient for any type of triangle pattern. However, as partially\footnote{\cite{MCAssey} compares alternative clustering coefficients for complete weighted graphs} noticed also in \cite{MCAssey}, this coefficient does not properly account for the strength of a node, resulting in a clustering coefficient too affected by weights.
	
	To overcome this issue, we propose a new local clustering coefficient for weighted and directed networks based on a generalization of the clustering coefficient developed in \cite{Barrat_2004}. On one hand, our proposal takes into account the triangles that the neighbours of a node $i$ form, completely preserving the initial idea of the clustering coefficient. On the other hand, the weights of these triangles also affect the coefficient. Following \cite{Barrat_2004}, in our proposed clustering coefficient we do not consider the weight of the closing link of a triangle (i.e. the link between adjacent neighbours of $i$). This is because the aim of the clustering coefficient is to assess the likelihood of the occurrence of this link that closes the triangle, and not its weight. A proper normalization of the local coefficient is assured by considering the strength of the node. Hence, both the number of triangles and their weights are captured by our coefficient, that in this way well replicate, for weighted and directed networks, the idea of nodes to be "clustered together". 
	
	Numerical results point out that the proposed coefficient proves effective in capturing both the number of closed triangles and the presence of strong neighbours (i.e. with higher weights), that classical indicators fail to correctly detect.
	The coefficient treats all possible directed triangles as they were the same, as if directions of edges were irrelevant. Furthermore, as in \cite{Fagiolo_2007}, we are able to provide alternative coefficients that only consider particular types of directed triangles. In other words, the proposed measure is capable to distinguish different patterns of directed triangles from a node perspective. In this way, we allow for different interpretation in terms of the resulting patterns.
	
	The paper is organized as follows. Section \ref{sec:Pr} introduces some basic definitions and notations used in the next; Section \ref{Clust_WN} briefly reviews some local clustering coefficients provided in the literature for weighted undirected network. Section \ref{clust_WDN} reports the coefficient given in \cite{Fagiolo_2007} and describes the alternative coefficient we propose. In Section \ref{sec:SC}, we look at directed networks at different \lq\lq observation scales\rq\rq in order to separately catch different patterns. Four clustering coefficients are here proposed, whose weighted average coincides with the overall coefficient.
	Section \ref{sec:ES} provides examples and numerical results, to compare our procedure to classical coefficients on the empirical networks considered. Section \ref{sec:conc} concludes.

	\section{Preliminaries}\label{sec:Pr}
	
	Formally, a directed graph (or digraph) $D=(V,A)$ is a pair of sets $V$ and $A$, where $V$ is the set of $n$ vertices (or nodes) and $A$ is the ordered set of $m$ pairs (arcs) of vertices of $V$; if $(i,j)$ or $(j,i)\in A$, then vertices $i$ and $j$ are adjacent. 
	
	A weight $w_{ij}>0$ can be associated with each link $(i,j)$ so that a weighted directed graph is obtained; we assume that, if omitted, the weight $w_{ij}$ of an arc $(i,j)$ is equal to 1 (binary case). In general, both adjacency relationships between vertices of $D$ and weights on the arcs are described by a nonnegative, real $n$-square matrix $\textbf{W}$ (the weighted adjacency matrix). In the unweighted case, matrix $\textbf{W}$ is simply the classical binary matrix $\textbf{A}$ (the adjacency matrix). In the next, we will consider the case of either unweighted or weighted graphs with no loops (i.e. $a_{ii}=0, w_{ii}=0$ $\forall i$).
	
	We call $G=(V,E)$ the graph in which every edge corresponds to an arc $(i,j)$ or $(j,i)$ in $D=(V,A)$. Observe that $G$ is a weighted graph and to every arc $(i,j)$ with weight $w_{ij}>0$ corresponds an edge $(i,j)$ with weights $w_{ij}=w_{ji}$. $G$ represents the "symmetric case" of $D$.
	The matrix $\textbf{W}$ (or $\textbf{A}$, for the unweighted case) associated to $G$ is, of course, a symmetric matrix. 
	The $(i,j)$ element of the $k-$power of the \textbf{A} is the number of walks of length $k$ from $i$ to $j$. 
	
	Using the same notation as in Fagiolo (\cite{Fagiolo_2007}), we define the in-degree (respectively out-degree) of a node $i$ as the number of arcs pointing towards (respectively starting from) $i$:
	\begin{equation}\label{in_deg}
		d^{in}_i=\sum_{j\neq{i}}a_{ji}=\textbf{A}^T_i\textbf{1}
	\end{equation}
	
	\begin{equation}\label{out_deg}
		d^{out}_i=\sum_{j\neq{i}}a_{ij}=\textbf{A}_i\textbf{1}.
	\end{equation}
	where $\textbf{A}_i$ and $\textbf{A}^T_i$ are respectively the $i-$th row of \textbf{A} and of its transpose, \textbf{1} is the unit column vector of $n$ elements. The degree   $d^{tot}_i$ of a vertex is then:
	\begin{equation}\label{tot_deg}
		d^{tot}_i=d^{in}_i+d^{out}_i=(\textbf{A}^T+\textbf{A})_i\textbf{1}.
	\end{equation}
	Bilateral arcs between  adjacent nodes $i$ and $j$, if any, are represented as:
	\begin{equation}\label{bil_deg}
		d^{\leftrightarrow}_i=\sum_{j\neq{i}}a_{ij}a_{ji}=\textbf{A}^{2}_{ii}.
	\end{equation}
	
	
	Moving to the weighted case, the previous definitions can be replaced by the strength of a node $i$:
	\begin{equation}\label{in_str}
		s^{in}_i=\sum_{j\neq{i}}a_{ji}w_{ji}=(\textbf{A}^T\textbf{W})_{ii}=\textbf{W}^T_i\textbf{1}
	\end{equation}
	
	\begin{equation}\label{out_str}
		s^{out}_i=\sum_{j\neq{i}}a_{ij}w_{ij}=(\textbf{AW}^T)_{ii}=\textbf{W}_i\textbf{1}.
	\end{equation}
	
	\noindent The total strength of $i$ is then:
	\begin{equation}\label{tot_str}
		s^{tot}_i=s^{in}_i+s^{out}_i=\sum_{j\not=i}\left(a_{ji}w_{ji}+a_{ij}w_{ij}\right)=(\textbf{A}^T\textbf{W}+\textbf{AW}^T)_{ii}=(\textbf{W}^T+\textbf{W})_i\textbf{1}.
	\end{equation}
	We define the strength related to bilateral arcs between adjacent nodes $i$ and $j$ as:
	\begin{equation}\label{bil_str}
		s^{\leftrightarrow}_i=\sum_{j\neq{i}}a_{ij}a_{ji}\frac{(w_{ij}+w_{ji})}{2}.
	\end{equation}
	
	Formula (\ref{bil_str}) can be also expressed in matrix form, as follow:
	\begin{equation}\label{bil_str1}
		\begin{split}
			s^{\leftrightarrow}_i=\sum_{j\neq{i}}a_{ij}a_{ji}\frac{(w_{ij}+w_{ji})}{2}=
			\frac{1}{2}\sum_{j\neq{i}}a_{ij}a_{ji}(w_{ij}+w_{ji})=\\
			\frac{1}{2}\sum_{j\neq{i}}(a_{ji}w_{ij}+a_{ij}w_{ji})=
			\frac{(\textbf{WA}+\textbf{AW})_{ii}}{2}.
		\end{split}
	\end{equation}
	
	\noindent where the second-to-last equality holds recalling that $w_{ij}\neq 0$ if and only if $a_{ij}=1$.
	
	\noindent Few comments about previous formulas are useful. \\
	Formula (\ref{bil_str}) extends (\ref{bil_deg}) to the weighted case, multiplying each bilateral link by the arithmetic mean of its weights\footnote{This is one possible choice for taking weights into account. Other choices are possible but could be not equally effectiveness in computation.}.
	The bilateral arcs contribute just once to the bilateral degree $d^{\leftrightarrow}_i$ of a node. Generalizing to $s^{\leftrightarrow}_i$ by means of formula (\ref{bil_str}), this fact still holds. In this case, the bilateral strength sums the average weight of each bilateral arc observed. Taking this in mind, in case of an undirected network, the degree and strength ($d_i$ and $s_i$) of $i$ are simply expressed as $d_i=d^{tot}_i-d^{\leftrightarrow}_i$ and $s_i=s^{tot}_i-s^{\leftrightarrow}_i$.
	
	
	\section{Clustering in weighted networks}\label{Clust_WN}
	Local clustering coefficient has been formalized in the paper of Watts and Strogatz (\cite{Watts_1998}). For a given node $i$, the clustering coefficient is the number of triangles $t(i)$ connected to this node divided by the number of triples (i.e. potential triangles) centered on it:
	\begin{equation}\label{Clust}
		C_i(\textbf{A})=\frac{2|t(i)|}{d_i(d_i-1)}=\frac{\textbf{A}^3_{ii}}{d_i(d_i-1)}
	\end{equation}
	where $\textbf{A}^3_{ii}$ counts twice the number of triangles in which a node $i$ participates.\\
	The average value $\bar{C}=\frac{1}{n}\sum_{i=1}^{n}C_i(\textbf{A})$ gives a global indicator of the network and it has been extensively used in the analysis of complex networks.
	
	Watts and Strogatz model works on undirected and unweighted networks. However, real networks are frequently weighted and several clustering coefficients have been properly designed to weighted, undirected networks in the literature (see for instance \cite{Holme2007}, \cite{Serrano2006} and \cite{Zhang2005}).\\
	A quite natural extension to the weighted case is provided by Onnela et al. (\cite{Onnela_2005}):
	\begin{equation}\label{Onn_Clust}
		C^{Onn}_{i}(\mathbf{\tilde{W}})=\frac{\sum_{j}\sum_{k\neq j}\tilde{w}_{ij}^{1/3}\tilde{w}_{jk}^{1/3}\tilde{w}_{ki}^{1/3}}{d_i(d_i-1)}=\frac{(\mathbf{\tilde{W}}^{\left[\frac{1}{3}\right]})^{3}_{ii}}{d_i(d_i-1)}
	\end{equation}
	where $\mathbf{\tilde{W}^{\left[\frac{1}{3}\right]}}=\left[\tilde{w}_{ij}^{\frac{1}{3}}\right]$, being $\tilde{w}_{i,j}=\frac{w_{i,j}}{max(w_{i,j})} \forall i,j$.
	Observe that in (\ref{Onn_Clust}), the total number of the triangles $t(i)$ is substituted by the geometric mean of the links' weights.\footnote{Being the network undirected, every arc $(j,k)$ appears twice in the formula.} 
	Main idea of this generalization is to replace the total number of the triangles in which a node $i$ participates, with the ``intensity'' of the
	triangle, defined here as the geometric mean of its weights.
	
	Barrat et al. (\cite{Barrat_2004}) proposes a different generalization:
	
	\begin{equation}\label{Barr_Clust}
		C^{Barr}_{i}(\textbf{W})=\frac{1}{s_i(d_i-1)}{\sum_{j}\sum_{k\neq j}\frac{w_{ij}+w_{ik}}{2}a_{ij}a_{ik}a_{jk}}.
	\end{equation}
	
	\noindent We rewrite formula (\ref{Barr_Clust}) in a more convenient form, using matrices $\textbf{A}$ and $\textbf{W}$ (recall that $w_{ij}\neq 0$ if and only if $a_{ij}=1$):
	\begin{equation}\label{Barr_Clust1}
		\begin{split}
			C^{Barr}_{i}(\textbf{W})&=\frac{1}{{s_i(d_i-1)}}\sum_{j}\sum_{k\neq j}\frac{w_{ij}+w_{ik}}{2}a_{ij}a_{ik}a_{jk} =
			\frac{1}{{s_i(d_i-1)}}\sum_{j}\sum_{k\neq j}w_{ij}a_{ij}a_{ik}a_{jk} =\\
			&  \frac{1}{{s_i(d_i-1)}}\sum_{j}\sum_{k\neq j}w_{ij}a_{ik}a_{jk} =
			\frac{(\textbf{W}\textbf{A}^{2})_{ii}}{s_i(d_i-1)}.
		\end{split}
	\end{equation}
	
	Notice that the number of triangles that appears in (\ref{Clust}) are replaced in (\ref{Onn_Clust}) and (\ref{Barr_Clust}) by the average of the weights of the links between the node $i$ and its neighbours $j$ and $k$. With respect to (\ref{Onn_Clust}), the definition, provided in \cite{Barrat_2004}, considers only two of the three link weights involved in a closed triangle, namely, those adjacent to node $i$ (i.e. $w_{i,j}$ and $w_{i,k}$). It requires that a link exist also between nodes $j$ and $k$ but does not take its weight $w_{j,k}$ into account. At the same time, weights' normalization is not needed in this case and the strength of the node $i$ is considered at the denominator. Finally, the geometric mean used in (\ref{Onn_Clust}) has been here replaced by arithmetic mean. This issue has been already analyzed in \cite{Opsahl}, where different methods for defining the triplet value (as geometric and arithmetic mean) have been compared with regard to the transitivity coefficient. Authors clearly state that arithmetic mean is less robust against extreme values but, at the same time, they show that the effect on the coefficient is often negligible.
	
	\noindent The previous formulas represent two different ways of generalization of the clustering idea and actually they provide slightly different results, also when applied to very simple graphs. The shape of weights link distribution obviously affects the differences between these approaches.

	\section{Clustering in weighted and directed networks}\label{clust_WDN}
	
	In this section we propose a new local clustering coefficient (see formula (\ref{our_clust})) for weighted and directed network. This proposal is based on a generalization of formula (\ref{Barr_Clust1}) to directed networks. The numerator of the coefficient takes into account all directed triangles that a node $i$ actually forms with its neighbours, weighted with the average weight of the links connecting a node $i$ to its adjacent $j$ and $k$. Then, it is divided by all possible (appropriately weighted) directed triangles that it could form\footnote{The coefficient is here defined using the matrix form to make the expression easier.}:
	
	\begin{equation}\label{our_clust}
		C^{*}_i(\textbf{W})=\frac{\frac{1}{2}[(\textbf{W}+\textbf{W}^T)(\textbf{A}+\textbf{A}^T)^2]_{ii}}{{s_{i}^{tot}\left( d_{i}^{tot}-1\right) -2s_{i}^{\leftrightarrow }}}
	\end{equation}

	\noindent As also stressed before, we sum, for all triplets formed in the neighbourhood of the vertex $i$, the average weight of the two participating edges of the vertex $i$. As in \cite{Barrat_2004}, the coefficient is only affected by the likelihood of the occurrence of the link between the adjacent of $i$, not by its weight. \\
	Furthermore, the denominator is properly arranged to consider the strength of the node. It accounts for the average weight of links incoming and outcoming from node $i$ times the maximum possible number of triplets in which the vertex may participate\footnote{We remind indeed that $s_{i}^{tot}=\bar{w}^{tot}_{i}d_{i}^{tot}$, where $\bar{w}^{tot}_{i}=\frac{\sum_{j\neq{i}}a_{ij}w_{ij}}{d_{i}^{tot}}$  is the average weight of links incoming and outcoming from node $i$.} and it ensures that $C^{*}_i(\textbf{W}) \in \left[0,1\right]$.
	\noindent It is noticeable that weighted bilateral arcs ($2s^\leftrightarrow_i$) have to be removed by the formula, as they represent "false" triangles, being formed by $i$ and by a pair of directed arcs pointing to the same node, e.g., $i \rightarrow j$ and $j \rightarrow i$. We have indeed that a node $i$ can be possibly linked to ${d_{i}^{tot}}\choose{2}$ pairs of neighbours. Being the network directed, a node $i$ can form up to two triangles with each pair, also including two "false" potential triangles for each bilateral link,
	also including two "false" potential triangles for each bilateral link. 
	
	Notice that if the network is undirected, $C^{*}_i(\textbf{W})= C^{Barr}_{i}(\textbf{W})$. Indeed being $\textbf{W}=\textbf{W}^T$, $\textbf{A}=\textbf{A}^T$ and $s^{\leftrightarrow}_i=s_i$ (by \ref{bil_str}) then $s^{tot}_i=2s_i$, it yields to: 
	
	\begin{equation}
		C^{*}_i(\textbf{W})=\frac{\frac{1}{2}[(2\textbf{W})(2\textbf{A})^2]_{ii}}{2{s_{i}\left( 2d_{i}-1\right) -2s_{i}}}=\frac{4[\textbf{W}\textbf{A}^2]_{ii}}{2{s_{i}\left( 2d_{i}-2\right)}}=C^{Barr}_{i}(\textbf{W})
	\end{equation}
	
	A comparison between formula (\ref{our_clust}) and the clustering coefficient for the binary case (provided in \cite{Fagiolo_2007}), gives additional information on the correlation between weights and topology. If a higher coefficient is observed in the weighted network, we are in presence of a network in which the triangles are more likely formed by the edges with larger weights. On the other hand, a lower value of (\ref{our_clust}) signals a network in which the topological clustering is generated by edges with low weight. In other words, the largest part of the interactions is occurring on edges not belonging to interconnected triplets. 
	
	A generalization of the clustering coefficient to weighted and directed networks has been previously studied in Fagiolo \cite{Fagiolo_2007} where the following coefficient, based on an extension of formula (\ref{Onn_Clust}), is proposed:
	
	\begin{equation}\label{clust_w_dir}
		C^{Fag}_i(\mathbf{\tilde{W}})=\frac{\frac{1}{2}[(\mathbf{\tilde{W}}^{\left[\frac{1}{3}\right]}+(\mathbf{\tilde{W}}^T)^{\left[\frac{1}{3}\right]}]^{3}_{ii}}{d^{tot}_i(d^{tot}_i-1)-2d^\leftrightarrow_i}
	\end{equation}
	
	\noindent It is worth mentioning that the number of actual weighted directed triangles connected to $i$ is divided by all its potential unweighted directed triangles. Even in this case, eventual bilateral edges ($2d^\leftrightarrow_i$) have to be removed. \\
	Notice that two main differences can be noted between formulas (\ref{our_clust}) and (\ref{clust_w_dir}). On one hand, weights, associated to edges in the neighbourhood of $i$, affect the coefficients in a different way. The total contribution of a weighted triangle is indeed defined in $C^{Fag}_i(\mathbf{\tilde{W}})$ as the geometric mean of the weights of all edges involved. Alternatively, the average weight of the two participating edges of the vertex $i$ is instead considered in formula (\ref{our_clust}). On the other hand, the denominator of the coefficient does not consider the node strength in $C^{Fag}_i(\mathbf{\tilde{W}})$, while it is considered by $C^{*}_i(\textbf{W})$. Indeed, $C^{Fag}_i(\mathbf{\tilde{W}})$ does not involve the actual strength of a node in the normalization factor, but only its maximum possible strength if all weights equal one (that is the node degree).  This often results in deflated clustering coefficients which cannot fully fulfill characteristics usually required to this kind of indicator. As the size of the network or the skewness of the weight link distribution increase, this deflation can become very serious.
	
	Also, it is easy to notice that both coefficients simply reduce to formula (\ref{Clust}) when $\mathbf{W}$ is binary and symmetric (unweighted and undirected network).
	
	Fagiolo also computes the expected value of the coefficient $E[C^{Fag}_i]$ for a weighted Erd\"os-R\'enyi $G(n,p)$ random graph. This specific weighted and directed random graph is obtained via a two-step algorithm: at first, the algorithm places each edge with independent probability $p \in(0,1)$ across all possible directed arcs. At the second step, it assigns the weights $w_{ij}$ of any existing directed arc sampling from an independent random variable uniformly distributed over the interval $(0,1]$. As a result, $E[C^{Fag}_i]=\left(\frac{3}{4}\right)^3p$.
	
	Recalling that a random graph evolves to the complete graph $K_{n}$ as $p$ approaches to 1, then $E[C^{Fag}_i]$  evolves to $\left(\frac{3}{4}\right)^3$ when the network tends to the complete directed network $K_{n}^{\leftrightarrow }$. Interestingly, our coefficient differently behaves when the network is complete, being equal to 1 regardless of the assigned weights (the computation of the coefficient for $K_{n}^{\leftrightarrow }$ is reported in Appendix A).
	
	\section{A focus on patterns of directed triangles}\label{sec:SC}
	
	As already pointed out in \cite{Fagiolo_2007}, in digraphs a node $i$ can be part of triangles with arcs that point in different directions, giving rise to completely different interpretation in terms of the resulting patterns.
	
	Following the same classification as in Fagiolo (see \cite{Fagiolo_2007}, Figure 1), we address to four types of triangles to which a node $i$ can take part (let $j$ and $k$ be the two other nodes involved).
	
	\begin{enumerate}
		\item \textit{In}, i.e. a triangle such that there are two arcs incoming into $i$ ($j \rightarrow i$, $k \rightarrow i$, $j \rightarrow k \vee k \rightarrow j$).
		\item \textit{Out}, i.e. a triangle such that there are two arcs coming out of $i$ ($i \rightarrow j$, $i \rightarrow k$, $j \rightarrow k \vee k \rightarrow j$).
		\item \textit{Cycle}, i.e. a triangle such that every arc has the same direction ($j \rightarrow i, i \rightarrow k, k \rightarrow j$ or vice versa).
		\item \textit{Middleman}, i.e. a triangle where the two arcs of $i$ have different directions and there is an arc between $j$ and $k$ (or vice versa), but without to form a cycle. In other words, there are two arcs incoming into $k$ or $j$ ($j \rightarrow i, i \rightarrow k, j \rightarrow k$ or vice versa).
	\end{enumerate}
	
	According to Fagiolo, we can specify a directed clustering coefficient for each one of the above cases, in order to have coefficients taking into account of the different patterns.
	Each coefficient is defined as the number of actual specific\footnote{In, out, cycle and Middleman.}  triangles of $i$ divided by the number of potential specific triangles of $i$, giving information about the clustering of the nodes respect to the pattern they are involved.
	
	A first component refers to triangles with two arcs in the node $i$ (\emph{in}-component):
	
	\begin{equation}\label{in clust}
		C^{*,in}_i(\textbf{W})=\frac{\sum_{j}\sum_{k\neq j}\frac{w_{ji}+w_{ki}}{2}a_{ji}a_{ki}\left(a_{jk}+a_{kj}\right)}{s_{i}^{in}\left( d_{i}^{in}-1\right)}=
		\frac{\frac{1}{2}[\textbf{W}^{T}(\textbf{A}+\textbf{A}^T)\textbf{A}]_{ii}}{s_{i}^{in}\left( d_{i}^{in}-1\right)}
	\end{equation}

	A second component regards triangles with two arcs starting from the node $i$ (\emph{out}-component):
	
	\begin{equation}\label{out clust}
		C^{*,out}_i(\textbf{W})=\frac{\sum_{j}\sum_{k\neq j}\frac{w_{ij}+w_{ik}}{2}a_{ij}a_{ik}\left(a_{jk}+a_{kj}\right)}{s_{i}^{out}\left( d_{i}^{out}-1\right)}=
		\frac{\frac{1}{2}[\textbf{W}(\textbf{A}+\textbf{A}^T)\textbf{A}^{T}]_{ii}}{s_{i}^{out}\left( d_{i}^{out}-1\right)}
	\end{equation}
	
	A third component considers middleman-type patterns (\emph{middleman}-component):
	
	\begin{equation}\label{midd clust}
		C^{*,middle}_i(\textbf{W})=\frac{\sum_{j}\sum_{k\neq j}\frac{w_{ji}+w_{ik}}{2}a_{ji}a_{ik}a_{jk}+\frac{w_{ij}+w_{ki}}{2}a_{ji}a_{ki}a_{kj}}
		{\frac{1}{2}\left(s_{i}^{in}d_{i}^{out}+s_{i}^{out}d_{i}^{in}\right)-s_{i}^{\leftrightarrow }}=
		\frac{\frac{1}{2}[\textbf{W}^{T}\textbf{AA}^{T}+\textbf{W}\textbf{A}^{T}\textbf{A}]_{ii}}{\frac{1}{2}\left(s_{i}^{in}d_{i}^{out}+s_{i}^{out}d_{i}^{in}\right)-s_{i}^{\leftrightarrow }}
	\end{equation}
	
	Finally, the last component refers to cyclical relation among $i$ and any two neighbours (\emph{cycle}-component):
	
	\begin{equation}\label{cycle}
		C^{*,cycle}_i(\textbf{W})=\frac{\sum_{j}\sum_{k\neq j}\frac{w_{ji}+w_{ik}}{2}a_{ji}a_{ik}a_{kj}+\frac{w_{ij}+w_{ki}}{2}a_{ij}a_{ki}a_{jk}}
		{\frac{1}{2}\left(s_{i}^{in}d_{i}^{out}+s_{i}^{out}d_{i}^{in}\right)-s_{i}^{\leftrightarrow }}
		=\frac{\frac{1}{2}[(\textbf{W}\textbf{A}^{2}+\textbf{W}^{T}(\textbf{A}^{T})^{2}]_{ii}}{\frac{1}{2}\left(s_{i}^{in}d_{i}^{out}+s_{i}^{out}d_{i}^{in}\right)-s_{i}^{\leftrightarrow }}
	\end{equation}
	
	As previously stated, each component conveys information about different patterns. Furthermore, the total number of actual directed triangles (i.e. the numerator of formula (\ref{our_clust})) can be split as the sum of the actual triangles considered by each component. In the same way, it is easy to prove that the maximum number of potential triangles (i.e. denominator of formula (\ref{our_clust})) can be obtained as the sum of the maximum number of cycle, middleman, in and out triangles that a single node can form (for a detailed proof of these results see Appendix B).
	
	Notice that, rearranging formula (\ref{our_clust}), we can re-express it as:
	\begin{equation}
		\label{Clust_WeightAv}
		\begin{split}
			C^{*}_i(\textbf{W})=C^{*,in}_i(\textbf{W})\frac{s_{i}^{in}\left( d_{i}^{in}-1\right)}{s_{i}^{tot}\left( d_{i}^{tot}-1\right) -2s_{i}^{\leftrightarrow}}
			+C^{*,out}_i(\textbf{W})\frac{s_{i}^{out}\left( d_{i}^{out}-1\right)}{s_{i}^{tot}\left( d_{i}^{tot}-1\right) -2s_{i}^{\leftrightarrow}}+\\
			C^{*,cycle}_i(\textbf{W}) \frac{\frac{1}{2}\left(s_{i}^{in}d_{i}^{out}+s_{i}^{out}d_{i}^{in}\right)-s_{i}^{\leftrightarrow}}{s_{i}^{tot}\left( d_{i}^{tot}-1\right) -2s_{i}^{\leftrightarrow}}
			+ C^{*,middle}_i(\textbf{W})
			\frac{\frac{1}{2}\left(s_{i}^{in}d_{i}^{out}+s_{i}^{out}d_{i}^{in}\right)-s_{i}^{\leftrightarrow}}{s_{i}^{tot}\left( d_{i}^{tot}-1\right) -2s_{i}^{\leftrightarrow}}
		\end{split}
	\end{equation}
	Hence the total clustering coefficient is a weighted average of the four coefficients ((\ref{out clust}), (\ref{in clust}), (\ref{midd clust}) and (\ref{cycle})), where weights are given by the denominator of each single component divided by the denominator of formula (\ref{our_clust}).
	
	The last expression is meaningful in capturing the relevance of the specific component within a single network. By formula (\ref{Clust_WeightAv}), it is evident that the values of both the coefficient and its weight have to be taken into account to assess the importance of each component.
	
	\section{Peculiarities of the two different coefficients}\label{sec:SC}
\begin{figure}[!h]
		\centering
		\includegraphics[width=12cm]{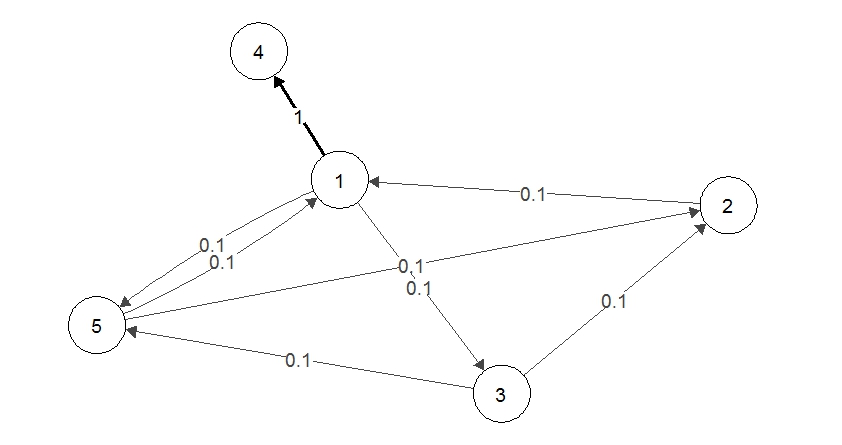}
		\caption{Graph $G_{3}$.}
		\label{fig:Example 2}
	\end{figure}

	We start comparing the behaviour of both local coefficients previously defined (see formulas (\ref{our_clust}) and (\ref{clust_w_dir})).\\ In order to better emphasize the main differences between these coefficients, it is convenient to focus on a simple digraph of order 5. The graph is weighted and all weights are equal to 0.1, except for the arc $(1,4)$, having weight 1 (see Figure \ref{fig:Example 2}). Neglecting the weights, the average total clustering coefficient ($\bar{C}^{Fag}(\mathbf{A})$), computed by Fagiolo's formula for BDN (see \cite{Fagiolo_2007}), is approximately 0.42. Extending this evaluation to the WDN, by using formula (\ref{our_clust}) the value is roughly 0.39 and this confirms a high-clustered network structure. We observe that $\bar{C}^{*}(\mathbf{W})$ is lower than $\bar{C}^{Fag}(\mathbf{A})$ being the topological clustering generated by edges with low weight.
	The coefficient is instead very close to zero (almost 0.05) by applying formula (\ref{clust_w_dir}) proposed in \cite{Fagiolo_2007}.
	
	It is worth noting that the coefficient of formula (\ref{clust_w_dir}) is more affected by weight values than the proportion between the number of actual triangles and the number of potential triangles. As the coefficient introduced in Onnela et al. for WUN (formula (\ref{Onn_Clust})), also the Fagiolo's coefficient (\ref{clust_w_dir}) does not involve the actual strength
	of a node, but only its maximum possible strength if all weights equal one.  Indeed, the actual triangles and the potential triangles are differently "weighted" in the coefficient's computation.
	
	This difference is noticeable not only at the global level, but also when its components (i.e. in, out, middle and cyclic) are separately inspected. For the sake of brevity, we only report in Table \ref{tab:clustExample_2} the cycle-type coefficient computed for graph $G_{3}$. Focusing for instance on vertex 2, it is worth pointing out that this node is involved in all potential cycles equally weighted (i.e. actual and potential cycles coincide with the same weights).
	Being only two cyclical relation ($2 \rightarrow 1 \rightarrow 5 \rightarrow 2$ and $2 \rightarrow 1 \rightarrow 3 \rightarrow 2$) possible between this vertex and any two of its neighbours, the coefficient is equal to one for the binary case (see \cite{Fagiolo_2007}). 
	Our coefficient well capture this behaviour, because all links have the same weight. Conversely, formula (\ref{clust_w_dir}) proves unable to catch this point in the right way, providing a very low coefficient (0.1).

	\begin{table}[!h]
		\small
		\centering
		\begin{tabular}{c | c | c | c | c | c |  }
			\hline
			Node	& Actual Cycles & Potential Cycles & $C^{Fag,cycle}_i(\mathbf{A})$  & $C^{Fag,cycle}_i(\mathbf{\tilde{W}})$  & $C^{*,cycle}_i(\textbf{W})$ \\
			\hline
			$1$ &3	&5	&0.6	&0.06 &0.21\\
			$2$ &2	&2  &1	&0.1 &1\\
			$3$ &2	&2  &1	&0.1 &1 \\
			$4$ &0	&0	&0	&0	&0\\
			$5$ &2 &3 &0.666	&0.066 &0.66\\
			$\bar{C}$ &	& &0.653	&0.065	&0.58\\
			\hline
		\end{tabular}
		\caption{Local and average Cyclic Clustering Coefficients for Graph reported in Figure \ref{fig:Example 2} }
		\label{tab:clustExample_2}
	\end{table}
	
\begin{figure}[!h]
		\centering
		\includegraphics[width=12cm]{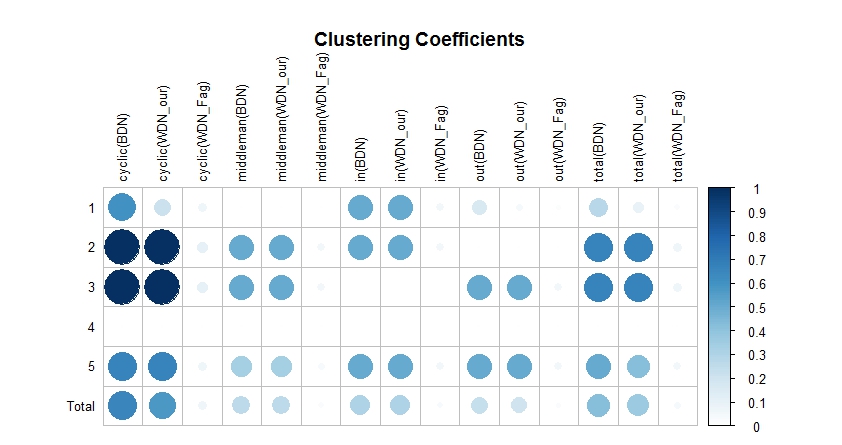}\\
		\caption{Local and average  clustering coefficients for graph $G_{3}$.}
		\label{fig:Example 2b}
	\end{figure}

	Figure \ref{fig:Example 2b} provides a summary view of alternative coefficients distinguishing different patterns of directed triangles. Rows provide local coefficients for each node, whereas the last row shows the average value. Columns report the different components for directed networks, weighted or not. The colour and the size of each bullet are different according to the coefficient's value.
	It can be observed how coefficient (\ref{our_clust}) (second column) mimics the behaviour of the coefficient applied for the binary case. As expected, greater differences are observed for nodes with a higher deviation between the average weights of the observed triangles and the strength (see node 1 for instance). On this example, formula (\ref{clust_w_dir}) seems unreliable leading to negligible coefficients in all cases.

	\section{Empirical analysis}\label{sec:ES}
	We now test the proposed clustering coefficient on a number of weighted and directed networks. The above concept can be meaningfully applied by considering several real networks belonging to different fields. To this aim we test the behaviour of formula (\ref{our_clust}) by assessing a systematic comparison with the existing coefficients provided in the literature. \\
	We begin by designing a global banking network using Bank for International Settlements' (BIS) consolidated statistics, which measure banks' exposures to different countries and sectors. They capture the worldwide consolidated claims of internationally active banks headquartered in BIS reporting countries. In particular, we consider international claims by a reporting country towards banks in counterparty countries. In this way, we focus on the lending activity of international banks. Here, nodes correspond to countries and weighted directed edges represent positive cross-border banking flows, i.e., increases in cross-border bank assets of a reporting country vis-\`{a}-vis another country. These are net flows in the sense that they account for repayments. For instance, a link regards investments (new loans, purchases of securities and other assets, etc.) of a country's banking system in another country minus repayments. 
	
	Data provided by BIS naturally induce a core/periphery structure, as the resulting network model is characterized by some nodes/countries densely-connected, whereas others are sparsely-connected in a peripheral position\footnote{ BIS indicates the list of countries included in the core. For this reason, core/periphery structure is not specifically referred to the classical definition of Borgatti and Everett (\cite{Borgatti2000}). However, it could be tested a priori the partition of nodes into core and periphery blocks, using the procedure in \cite{Borgatti2000}.}. It is noteworthy that in the core/periphery network the links (from the core to the periphery nodes) are only unidirectional, because periphery countries do not report data to the BIS. 
	
\begin{figure}[!h]
		\centering
		\includegraphics[width=6.5cm]{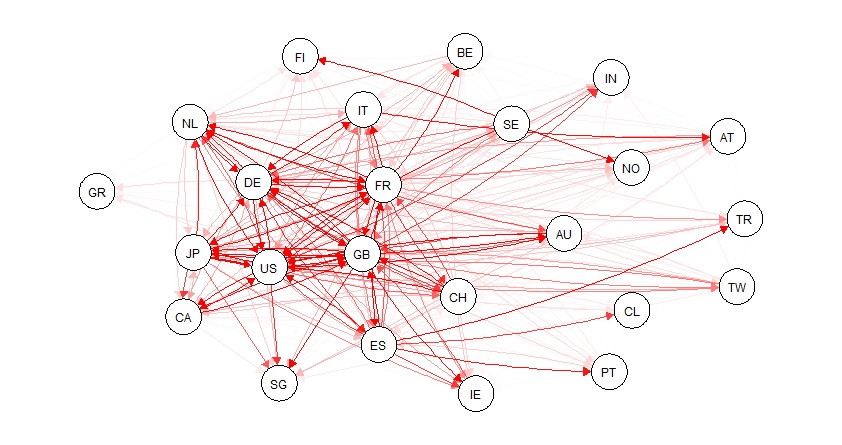}
		\qquad\qquad
		\includegraphics[width=6.5cm]{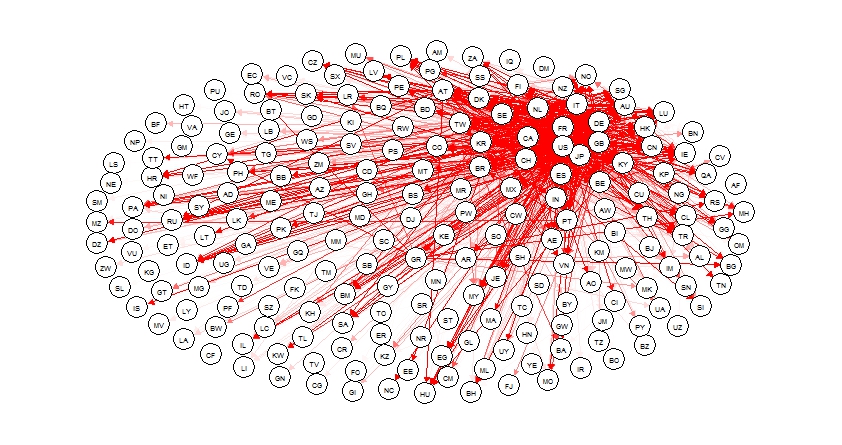}\\
		\caption{Cross-border global banking networks (Core/Core and Core/Periphery respectively). }
		\label{fig:BanksGraphs}
	\end{figure}

	We model each quarter of the year over the sample period (from first quarter of 2005 to third quarter of 2016) as a separate network and we analyse two different types of networks: the full one (core/periphery), which refers to links between banks of approximately\footnote{Number of countries varies according to the time because few isolated nodes are present at some specific time-periods.} 212 countries and the core/core network, which refers to links among the 24 core countries.  
	We report in Figure \ref{fig:BanksGraphs} both networks evaluated at the third quarter of 2016. Core/core network is very dense (density is approximately 0.85) with 24 vertices and 462 edges (this network would be a complete graph with 552 edges). We observe instead that density falls at 0.05 when periphery countries are considered too. On one hand, many nodes in the core lend to each other, on the other hand periphery countries are not related to each other because of data structure.
	
\begin{figure}[!h]
		\centering
		\includegraphics[width=6.5cm]{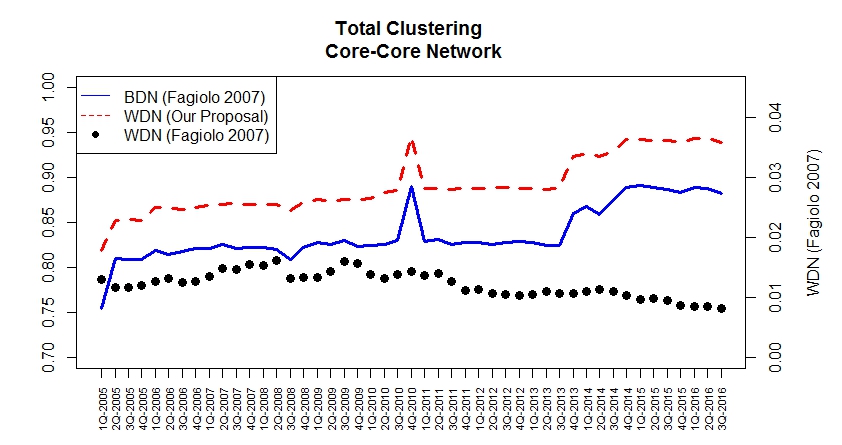}
		\qquad\qquad
		\includegraphics[width=6.5cm]{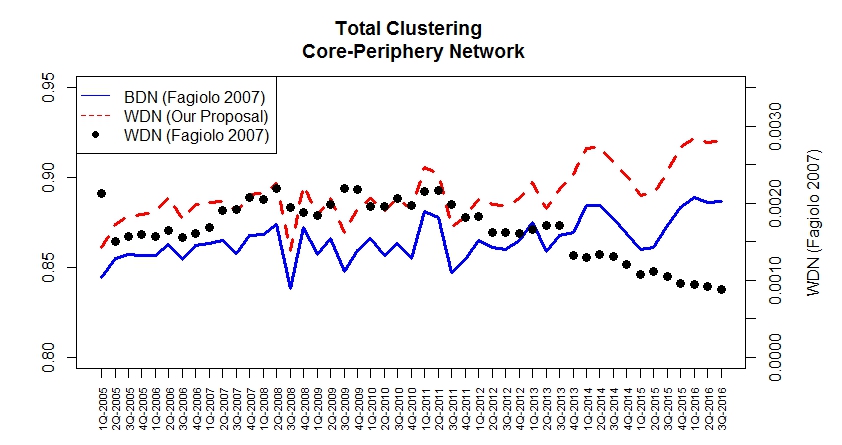}\\
		\caption{Binary and weighted directed total clustering coefficients. Figure reports $\bar{C}^{Fag}(\mathbf{A})$ (proposed by Fagiolo \cite{Fagiolo_2007} for BDN) and  $\bar{C}^{*}(\textbf{W})$ (our proposal for WDN). $\bar{C}^{Fag}(\mathbf{\tilde{W}})$,  proposed by Fagiolo (\cite{Fagiolo_2007}) for WDN, is assigned to the secondary scale ($y$-axis on the right side).}
		\label{fig:TotalClust}
	\end{figure}

	However, our aim is to focus on clustering coefficients. It is noteworthy that several works, contributing to the debate around systemic risk, showed (see \cite{Billio}, \cite{Minoiu} and \cite{Tabak}) that the directed clustering coefficient could provide meaningful insights in this context. In particular, in \cite{Tabak} the authors argue that higher clustering of the ``in'' type may reflect higher systemic risk because failure of the borrowing node in an ``in'' triangle can trigger simultaneous non-repayments to the lending nodes, and this, in turn, can make them unable to honor their own obligations. The implication of high clustering of the ``cycle'' variety is more ambiguous, since nodes in a ``cycle'' triangle act as both borrowers and lenders in the interbank market, so the consequences of a node failure are unclear. \\
	To this aim, we start reporting in Figure \ref{fig:TotalClust} patterns of total clustering coefficients computed by applying formula provided in \cite{Fagiolo_2007} for binary directed networks (BDN) and formulas (\ref{our_clust}) and (\ref{clust_w_dir}) for weighted directed networks (WDN). All coefficients have been determined by using both core/core and core/periphery networks. In the binary case, we observe high total clustering coefficients. Within the core, this is easily explained by the very high density. For instance, local clustering coefficients move from 0.85 to 0.92 in the last time period. Core/periphery network is also characterized by a significant clustering coefficient, but, in this case, the high value is a result of opposite behaviours between core and periphery countries. The former ones have now a lower coefficient because they are related to more than one periphery country (not connected each other because of the dataset structure). On the other hand, each periphery country shows a very high local clustering coefficient (both ``in'' and total type) borrowing money from at least two core countries that are usually connected each other. \\
	
\begin{figure}[!h]
		\centering
		\includegraphics[width=5cm]{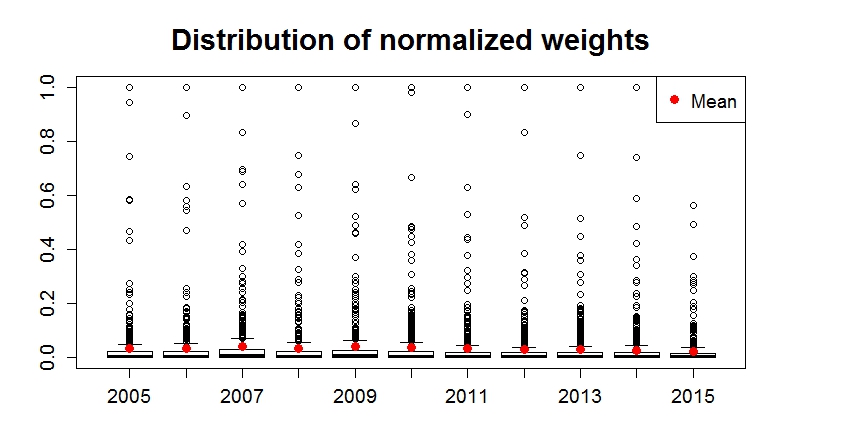}
		\caption{Distribution of Normalized Weights (Core-Core Networks)}
		\label{fig:NormWeight}
	\end{figure}

\begin{figure}[!h]
		\centering
		\includegraphics[width=6cm,height=6cm]{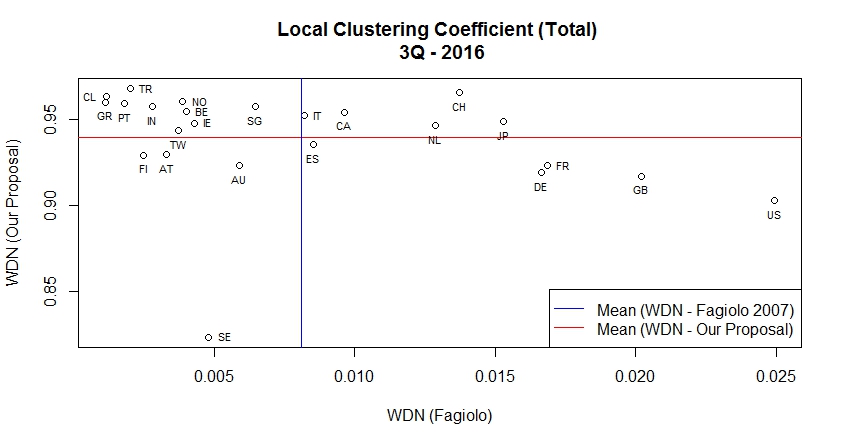}
		\qquad\qquad
		\includegraphics[width=6cm,height=6cm]{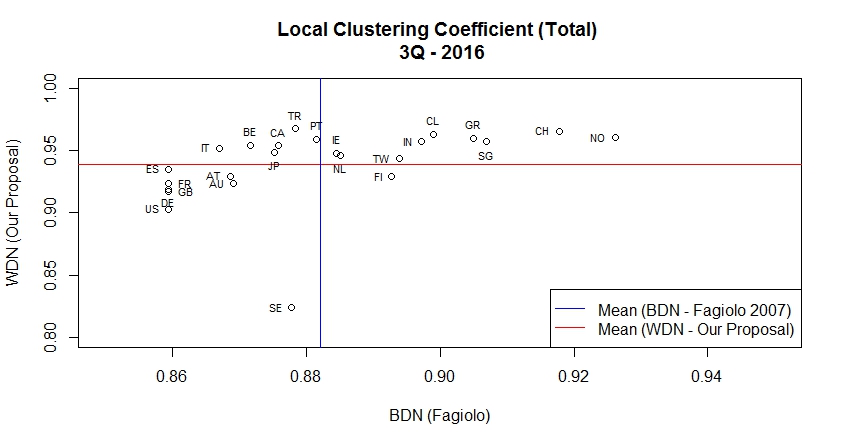} \\
		\includegraphics[width=6cm,height=6cm]{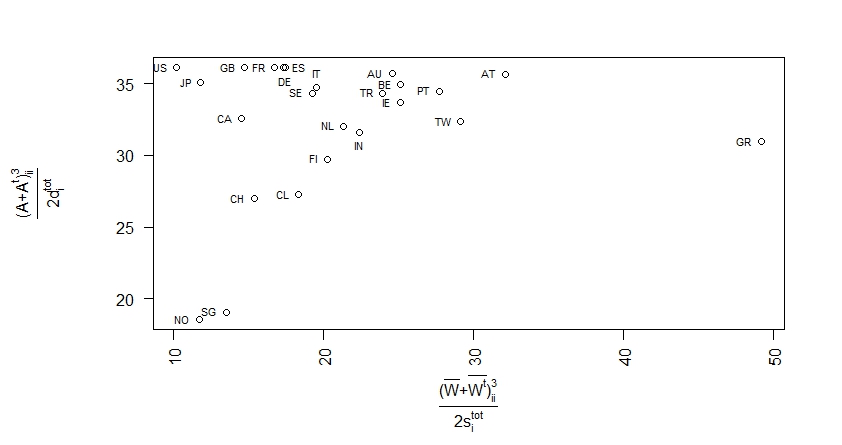}
		\qquad\qquad
		\includegraphics[width=6cm,height=6cm]{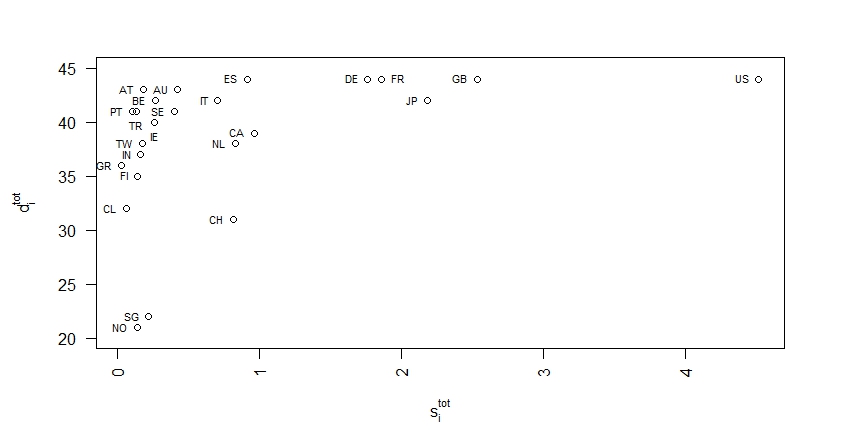}
		\caption{Upper Figures (a and b) provide a comparative illustration of alternative local clustering coefficients. They have been computed for weighted directed networks (WDN) by using formulae (\ref{clust_w_dir}) and (\ref{our_clust}), while for binary directed network (BDN) as described in \cite{Fagiolo_2007}. Lower figure on the left side (c) reports the ratio of the total contribution of observed triangles (defined as the geometric mean of its weights) to the strength of the node. It is compared to the ratio of number of triangles with the node as one vertex to the degree of the node. On the right side (d) the total degree and the total strength of each node is reported.}
		\label{fig:LocalCoefficients}
	\end{figure}

	Moving to the weighted case, we computed both formulas (\ref{our_clust}) and (\ref{clust_w_dir}). Weights have been normalized in the unit interval $[0,1]$ by dividing by the maximum observed flow, as proposed in \cite{Fagiolo_2007}. This operation has no impact on our proposal, but it assures that $C^{Fag}_i(\mathbf{\tilde{W}})$ (see (\ref{clust_w_dir})) ranges in $[0,1]$ too. \\
	As shown in Figure \ref{fig:TotalClust}, the overall average coefficient $\bar{C}^{Fag}(\mathbf{\tilde{W}})$ is very low. On the contrary, our extension of Barrat coefficient depicts a strongly clustered network when weights are considered. This result is also in line with weighted directed transitivity coefficient proposed by \cite{Opsahl}, that shows values very close to 1. \\
	We notice that $C^{Fag}_i(\mathbf{\tilde{W}})$ appears strongly affected by the weights' values involved in the observed triangles disregarding the strength of each node. In particular, the low average and the high skewness (equal to 0.006 and 17 respectively at the third quarter of 2016) of weight link distribution lead to very low local clustering coefficients (see Figure \ref{fig:NormWeight} for weight link distribution in core-core networks). To give an idea, at the third quarter of 2016 (3Q -2016), the $99^{th}$ percentile of weight link distribution is roughly 0.107 and only two arcs (JP$\rightarrow$US and CA$\rightarrow$US) have a weight greater than 0.5. \\
	Going deep into the analysis, it is noteworthy that countries with a higher average weight (i.e. US, GB, FR, JP, DE, CA, CH in decreasing order) show a greater clustering when formula (\ref{clust_w_dir}) is applied (see Figure \ref{fig:LocalCoefficients}a). But, at the same time, many of these countries (see US for instance) have also a significant strength and a lower ratio of weighted observed triangles to total strength (see Figure \ref{fig:LocalCoefficients}c). These results confirm that strength of the node is not properly taken into account in $C^{Fag}_i(\mathbf{\tilde{W}})$.  For the same countries, a clustering coefficient (see Figure \ref{fig:LocalCoefficients}b) lower than the binary one is instead observed by applying formula (\ref{our_clust}) because the average weights involved in the observed triangles are lower than the average weight of ``in-'' and ``out-flows'' of that node. \\
	On the other hand, we can emphasize the case of the Greece (GR) characterized by a prominent ratio of weighted observed triangles to total strength. In this case, $C^{*}_i(\mathbf{A})$ is higher than the average clustering, while  $C^{Fag}_i(\mathbf{\tilde{W}})$ is significantly lower.

\begin{figure}[!h]
		\centering
		\includegraphics[width=6cm,height=6cm]{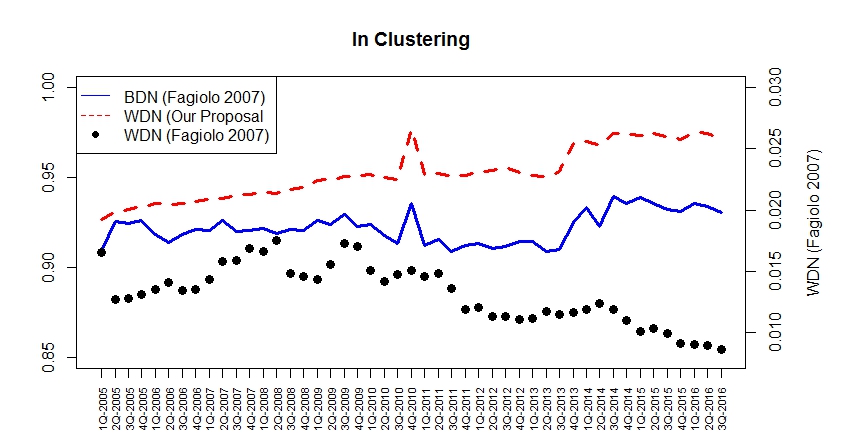}
		\qquad\qquad
		\includegraphics[width=6cm,height=6cm]{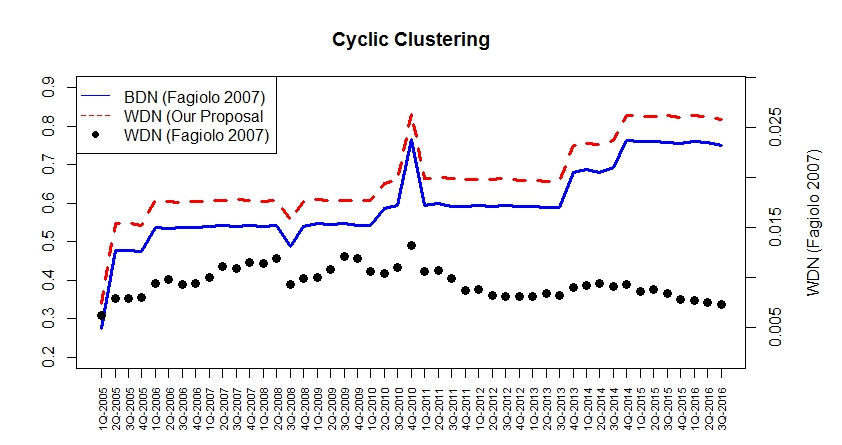}
		\caption{Average ``in'' and ``cycle'' clustering coefficients in both WDN and BDN cases (Core/Core network). Notice that $\bar{C}^{Fag,in}(\mathbf{\tilde{W}})$ and $\bar{C}^{Fag,cycle}(\mathbf{\tilde{W}})$ are assigned to the secondary scales ($y$-axis on the right side).}
		\label{fig:InCyclicClust}
	\end{figure}

	In order to disentangle the effect of specific triangle patterns (cycle, middleman, in and out), the four alternative directed clustering coefficients have been also exploited with regard to both networks. For the sake of brevity, we only report patterns of in and cycle variety computed for core/core network (see Figure \ref{fig:InCyclicClust}). \\
	Binary clustering coefficients associated to different triangle patterns show a relevant heterogeneity moving in the range (0.27 , 0.91) at first quarter of 2005 and in the interval (0.75, 0.93) at third quarter of 2016. Formulas (\ref{in clust}), (\ref{out clust}), (\ref{midd clust}) and (\ref{cycle}) have a similar behaviour ranging in (0.34,0.92) and (0.82,0.97) intervals respectively. Heterogeneity is caught also by coefficients proposed in \cite{Fagiolo_2007} but very low values have been observed (between 0.007 and 0.009 at 3Q-2016). \\
	
	Futhermore, in-clustering is relatively high and quite stable over the long run. It reached a peak at the end of 2010, probably as a result of the financial crisis, and it showed a slow increasing tendency after 2012.
	According to the cycle component, we observe a significant increase in the number of cyclical relations over time. Although this behaviour is probably justified by a general tendency to diversify the relationships (also explained by higher in- and out-degree), it could be interesting to analyse how fast a cascade effect could propagate in a so interconnected system. \\
	
	\newpage
	We now focus on a second empirical application based on a U.S. airport network. This network is a modified version of U.S. airport network used in \cite{Colizza}, \cite{OpsahlNEt} and \cite{Opsahl} and based on 2002 data of 500 busiest commercial airports. We used the Air Carrier Statistics database (available on the US Department of Transportation), also known as the T-100 data bank, that contains domestic and international airline market and segment data. Both certificated U.S. air carriers and foreign carriers (having at least one point of service in the United States or one of its territories) report monthly air carrier traffic information using specific forms. The data is collected by the Office of Airline Information, Bureau of Transportation Statistics, Research and Innovative Technology Administration. \\
	In the network we built, two airports are connected if a flight was scheduled between them in a given year. We analysed separately data from 2014 to 2016\footnote{At moment, data of 2016 are only available until the end of September}. The weight of a tie corresponds to the number of enplaned passengers\footnote{Number of enplaned passengers is the most important air traffic metric, because the majority of airport revenues are generated directly or indirectly from enplaned passengers. The term enplaned passenger is widely used in the aviation industry and it is loosely defined as a passenger boarding a plane at a particular airport. Data considers the total number of revenue passengers boarding an aircraft (including originating, stopover, and transfer passengers) in scheduled and non-scheduled services}. It considers revenue enplaned passengers within the U.S. as well as passengers enplaned outside U.S. but deplaned within the U.S\footnote{In \cite{Opsahl} the authors used instead only domestic data and weights equal to the number of seats available on the scheduled flights.}.  \\
	The airport network used in literature is highly symmetric so that it is usually analysed as an undirected one (see \cite{Barrat_2004} and \cite{Opsahl}). In our case, we observe a strong correlation (close to 1) between in- and out-degree  (or between in and out-strength in the weighted case). 
	The (non-scaled) $S$ measure, proposed in \cite{Fagiolo_2006} to assess whether an empirically-observed network is sufficiently symmetric to justify an undirected network analysis, is not too far from zero (equal to 0.19 and 0.02 in BDN and WDN cases respectively). Hence, network is then weakly asymmetric with a more pronounced behaviour in the binary case. Furthermore, we obtain a bigger network than that used in \cite{Opsahl} based on roughly 1600 airport and 25000 edges. \\
	In 2014 network, on average, each airport is connected to roughly 16 other airports (i.e., density is 0.01).
	For the average route, roughly 500 thousand enplaned passengers were observed.
	In this network, the average clustering coefficient is relatively high in both binary and weighted case ($\bar{C}^{*}(\mathbf{W})$) (see Table \ref{tab:AirportRes}). On the other hand, $\bar{C}^{Fag}(\mathbf{\tilde{W}})$ shows values close to zero. Furthermore, it is worth mentioning that directed transitivity \cite{Opsahl} provides values close to these results (equal to 0.38 and 0.6 for BDN and WDN in 2014). Instead, treating the network as an undirected one, it had turned out a little bit different in the binary case. The average clustering coefficient is indeed equal to 0.21 in 2014. Also a slight heterogeneity is observed by considering specific clustering coefficients based on different triangle patterns. \\
	Furthermore, the network is characterized by a giant component (GC). The presence of a giant component is a common issue in networks from real data and most papers focus on it.  Clustering coefficients evaluated on the GC of 2014 network, are equal to 0.54 and 0.64 with $\bar{C}^{Fag}(\mathbf{A})$ and $\bar{C}^{*}(\mathbf{W})$ respectively, while the average clustering coefficient based on (\ref{clust_w_dir}) shows the same value observed on the overall network (i.e. approximately 0.004). \\
	Similar results have been observed also in 2015 and 2016, where average clustering coefficients very close to directed transitivity coefficients (equal to 0.37 and 0.35 for BDN in 2015 and 2016 and equal to 0.64 and 0.61 for WDN in 2015 and 2016) are obtained.
	
	\begin{table}[!h]
		\begin{tabular}{l || c | c |c || c | c | c|| c| c| c|| }
			\hline
			&\multicolumn{3}{c||}{\textbf{Airport Network (2014)}}&\multicolumn{3}{c||}{\textbf{Airport Network (2015)}}&\multicolumn{3}{c||}{\textbf{Airport Network (2016)}} \\	
			\hline
			& $\bar{C}^{Fag}(\mathbf{A})$  & $\bar{C}^{Fag}(\mathbf{\tilde{W}})$ & $\bar{C}^{*}(\mathbf{W})$ &  $\bar{C}^{Fag}(\mathbf{A})$  & $\bar{C}^{Fag}(\mathbf{\tilde{W}})$ & $\bar{C}^{*}(\mathbf{W})$ &  $\bar{C}^{Fag}(\mathbf{A})$  & $\bar{C}^{Fag}(\mathbf{\tilde{W}})$ & $\bar{C}^{*}(\mathbf{W})$\\	\hline
			cyclic&  0.4773	&0.0031	&0.5632 			&0.4411	&0.0029	&0.5192&0.4366	&0.0040 &0.5141\\
			middleman&0.4731	&0.0031	&0.5587 			&0.4350	&0.0029	&0.5133			&0.4294	&0.0040	&0.5064\\
			in&0.4565	&0.0032	&0.5381 			&0.4078	&0.0029	&0.4821			&0.4090	&0.0042	&0.4801\\
			out&0.4438 &	0.0033	&0.5246 		&0.4120	&0.0030	&0.4863			&0.4089	&0.0041	&0.4831\\
			total&0.4852	&0.0031	&0.5709 			&0.4453	&0.0029	&0.5235			&0.4405	&0.0040	&0.5178\\ \hline
		\end{tabular}
		\caption{Comparison between average clustering coefficients (U.S. Airport Networks 2014-2016)}
		\label{tab:AirportRes}
	\end{table}
	
\begin{figure}[!h]
		\centering
		\includegraphics[width=6cm,height=6cm]{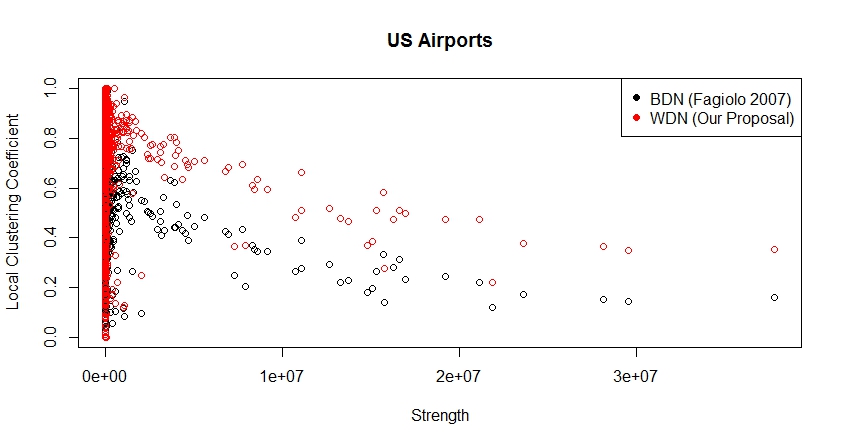}
		\qquad\qquad
		\includegraphics[width=6cm,height=6cm]{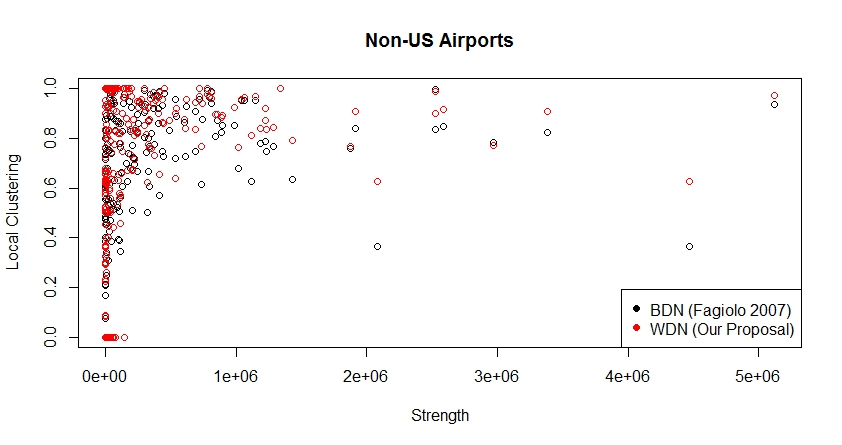}
		\caption{Local Clustering Coefficients vs out-strength for US and Non-US Airports (BDN and WDN US-Airport Network 2016)}
		\label{fig:LocalClustAirp}
	\end{figure}

	Local Clustering Coefficients can provide meaningful insights. In particular, Figure \ref{fig:LocalClustAirp} reports patterns of local clustering coefficients and out-strength differentiated between US and Non-US airports. It is noticeable that airports with busy routes are part of triplets, showing a greater clustering coefficient when number of passengers are taken into account (i.e. $C^{*}(\mathbf{W})\geq C^{Fag}(\mathbf{A})$ in many cases). We are indeed in presence of a network in which the interconnected triplets are more likely formed by the arcs with larger weights.\\
	Moreover, a higher local clustering coefficient is on average observed for Non-US airports. Usually, these airports tend to be linked to busier US airports (on average each one is related to 6 airports) that are usually connected each other. We remind indeed that edges between Non-US airports are not considered in this network. Furthermore, we observe a strong correlation between strength and local clustering (equal to 0.27 and 0.29 in BDN and WDN networks). Airports that are more strongly connected tend to form more strongly connected trade circles.\\
	According to US Airport, although the low correlation between local clustering and out-strength (-0.07 and 0.03 in BDN and WDN respectively), it is noteworthy that very large airports show a lower local clustering. Higher out-degree of these airports also implies connections to peripheral and remote airports that are not part of a triangle.
	
	As shown in \cite{Fagiolo_2007}, the concepts, described in previous Sections, can be meaningfully illustrated in the case of the empirical network describing world trade among countries. At this regard, we exploit a third dataset based on world trade connections. Source data provide, for any given year, imports (and exports) from (and to) a large sample of countries. As in \cite{Fagiolo_2007}, we focus on the year 2000 only and we build an edge between any two countries if there is a non-zero trade between them. Furthermore, in order to consider natural correlation between exporting levels and ``size'' of countries, we compute weights as the ratio of exports' amount to the GDP of the country\footnote{Both exports' amount and GDP are expressed in 2000 US dollars. GDP has been obtained by World Bank data based on World Bank national accounts data and OECD National Accounts data files. According to the network, although the link provided in \cite{Fagiolo_2007} is not anymore available, we used the World Trade data used in \cite{SubWei}. We cannot assure that the network is exactly the same as that used in \cite{Fagiolo_2007}, but the results obtained in terms of clustering coefficients are very close.} as proposed in \cite{Fagiolo_2007}. Main results are reported in Table \ref{tab:WTN}.
	
	\begin{table}[!h]
		\centering
		\begin{tabular}{l | c | c |c | }
			\hline
			& $\bar{C}^{Fag}(\mathbf{A})$  & $\bar{C}^{Fag}(\mathbf{\tilde{W}})$ & $\bar{C}^{*}(\mathbf{W})$ \\	\hline
			cyclic& 0.7349	&0.0004	&0.8068\\
			middleman&0.7426	&0.0008	&0.8112\\
			in&0.8199	&0.0005	&0.8796\\
			out&0.7347	&0.0013	&0.8326\\
			total&0.8142	&0.0007	&0.8880\\ \hline
		\end{tabular}
		\caption{Comparison between average clustering coefficients (World Trade Network - 2000)}
		\label{tab:WTN}
	\end{table}
	
\begin{figure}[!h]
		\centering
		\includegraphics[width=8cm,height=6cm]{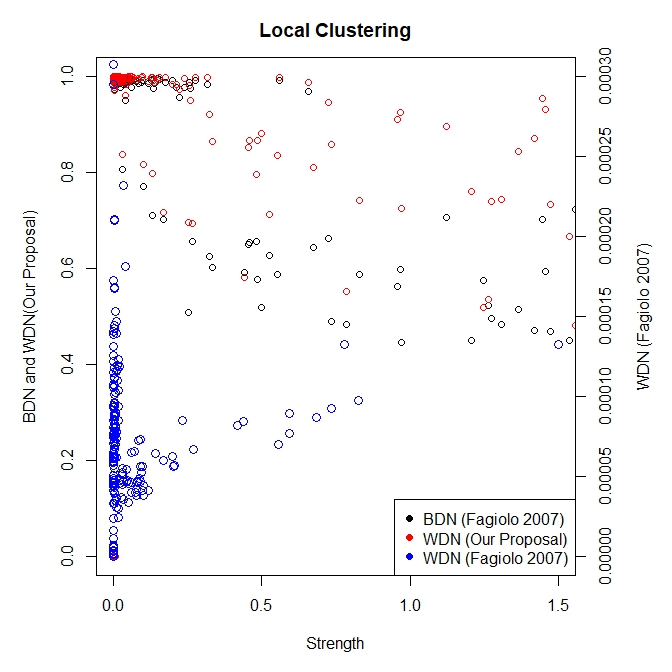}
		\caption{Local Clustering Coefficients vs total strength for World Trade Network. Notice that $C_{i}^{Fag}(\mathbf{\tilde{W}})$ is assigned to the secondary scale ($y$-axis on the right side).}
		\label{fig:LocalClustering_WTN}
	\end{figure}

	By using formula (\ref{our_clust}), we confirm the heterogeneity between weighted clustering coefficients as already shown in \cite{Fagiolo_2007}. But, also on this network, we derive results very far from $\bar{C}^{Fag}(\mathbf{\tilde{W}})$ and closer to the binary case. \\
	According to the relation between local clustering coefficients and total strength (see Figure \ref{fig:LocalClustering_WTN}), we observe a negative correlation in the binary network and a positive correlation when weights are introduced. Countries that are more strongly connected tend to form more strongly connected trade circles. In this context, we should expect an increase of weighted clustering coefficient with respect to the binary one. This results is indeed observed when formula (\ref{our_clust}) is applied, while a strong reduction is provided by (\ref{clust_w_dir}).
	
	Finally, we test our coefficient on the datasets analysed in \cite{Opsahl}. For the sake of brevity, we report only the overall coefficients in Table \ref{tab:OpsNet}.
	The first three networks are Freeman's EIES networks (see \cite{Freeman} and \cite{WasFaust}). This dataset regards three networks of researchers working on social network analysis. The first is an acquaintance network including 46 researchers, and in which relationships were recorded at the beginning of the study (time 1). Twelve researchers are isolated nodes and we removed them. The second network is similar, but the data were recorded at the end of the study (time 2). Also in this case twelve nodes are isolated. The third is a frequency matrix of the number of messages sent among 32 of the researchers that used an electronic communication tool.  In the first two networks, weights assume values between 0 and 4\footnote{A value equal to 4 represents a close personal friend of the researcher's; 3 represents a friend; 2 represents a person the researcher has met; 1 represents a person the researcher has heard of, but not met; and 0 represents a person unknown to the researcher.}. Both networks exhibit a large tendency toward clustering and our results are fully in line with the transitivity coefficient proposed in \cite{Opsahl}. We observe noticeable but lower differences with $\bar{C}^{D,Fag}_{i}(W)$, because of a lower skewness of weight link distribution. According to the Freeman EIES (messages) network, as detailed shown in \cite{Opsahl}, stronger ties are more likely to be part of triangles than weaker ties. This behaviour is also caught by $\bar{C}^{*}_{i}(W)$. \\
	Then, we analyse four organizational networks based on a consulting company ($n=46$) and a research team ($n=77$) in a manufacturing company \cite{CrossParker}. The ties in the consulting network are differentiated in terms of frequency of information or advice requests, whereas the ties in the Research Team network are differentiated in terms of the value placed on the information or advice received.  In these networks, ties are weighted on a scale from 0 to 5 and from 0 to 6 respectively. As already stressed in \cite{Opsahl}, data collection took place after an organizational restructuring operation that combined four separate units in different European countries. The research team was partitioned into strong communities based on the employees' previous geographical location (\cite{CrossParker}) and these reorganization have been partly responsible for a high value of clustering (\cite{Feld}). In these networks, according to both our proposal and clustering coefficient proposed in \cite{Opsahl}, weights have not a significant effect on clustering. Lower coefficients have instead been observed by applying (\ref{clust_w_dir}).\\
	Last network is the well-known neural network of the \emph{Caenorhabditis elegans} worm, studied in \cite{Watts_1998} and \cite{Achacoso}. \emph{Caenorhabditis elegans} is a small, free-living soil nematode (roundworm) that lives in many parts of the world and survives by feeding on microbes, primarily bacteria. The network\footnote{We used the \emph{Caenorhabditis elegans} network provided in t-net package of R. Number of nodes are equal to the network used in \cite{Opsahl} and greater than the network defined in \cite{Watts_1998}.} contains 297 (not isolated) nodes that represent neurons. A tie joins two neurons if they are connected by either a synapse or a gap junction. Weights represent the number of these synapses and gap junctions. It is a directed network with a low density (0.026). As shown in \cite{Watts_1998}, it could be considered a small-world network having a clustering greater than random and an average path close to the random graph.
	Also in this case, formula (\ref{clust_w_dir}) shows some inconsistencies providing a clustering coefficient close to zero.
	
	\begin{table}[!h]
		\centering
		\begin{tabular}{l | c| c | c |c | c | }
			\hline
			&n & $\bar{C}^{Fag}(\mathbf{A})$  & $\bar{C}^{Fag}(\mathbf{\tilde{W}})$ & $\bar{C}^{*}(\mathbf{W})$& $C_{\omega,am}$ \cite{Opsahl} \\	\hline
			
			Freeman EIES (time 1)	& 34 &0.77	&0.38	&0.78	&0.77\\
			Freeman EIES (time 2)	& 34 &0.82	&0.44	&0.83	&0.82\\
			Freeman EIES (messages)	& 32 &0.76	&0.03	&0.85	&0.74\\
			Consulting (advice)		& 46 &0.69	&0.29	&0.71	&0.71\\
			Consulting (value)		& 46 &0.68	&0.52	&0.69	&0.69\\
			Research team (advice)	& 77 &0.70	&0.31	&0.74	&0.71\\
			Research team (awareness) & 77	&0.71	&0.53	&0.73	&0.70\\
			C.\emph{elegans}' neural network & 297	&0.17	&0.01	&0.19	&0.24\\ \hline
		\end{tabular}
		\caption{Comparison between average clustering coefficients and transitivity coefficients}
		\label{tab:OpsNet}
	\end{table}
	
	\section{Conclusions}\label{sec:conc}
	
	A fundamental measure that has long gained attention in both theoretical and empirical research is the clustering coefficient. In particular, despite some well-known limitations, a local version is often needed in order to assign a specific measure of interconnection to each node. However, the problem of measuring and assessing local clustering for weighted directed network deserves further attention, as existing coefficients are not so effective in capturing interconnectedness when asymmetric weighted connections are observed. The problem is that the positive skewness of weight link distribution can lead to useless clustering coefficients being very close to zero. By extending the concept of clustering proposed by Barrat et al. \cite{Barrat_2004} for weighted undirected network, we provide a novel coefficient. In order to take edge directionality fully into account, we have also defined specific coefficients for each particular directed triangle pattern. Interestingly, we are capable to catch both the general clustering features of the underlying network and the relevance of each specific pattern.
	The approach has been tested on several real networks, clearly showing its advantages over classical coefficients.
	
	\section*{Acknowledgement}
	The authors are grateful to Anna Torriero and Gabriele Tedeschi for useful advice and suggestions.

	\appendix
	\section{Local clustering for directed complete network $K_{n}^{\leftrightarrow }$}
	
	We compute the clustering value of the $i$-th vertex $C_i(\textbf{W})$ for a directed complete network of order $n$, $K_{n}^{\leftrightarrow }$.
	It can be useful recalling that a digraph is said complete if $\forall u,v\in V$, both $(u,v)$ and $(v,u)$ belong to $A$
	(see \cite{Bang-Jensen2008}).
	By the formula (\ref{our_clust}), expanding the numerator, we have:
	\begin{eqnarray*}
		\begin{split}
			\frac{1}{2}[(\textbf{W}+\textbf{W}^T)(\textbf{A}+\textbf{A}^T)^2]_{ii}& =
			\sum_{j}\sum_{k\neq j}\frac{w_{ij}+w_{ik}}{2}a_{ij}a_{ik}\left(a_{jk}+a_{kj}\right) +
			\sum_{j}\sum_{k\neq j}\frac{w_{ji}+w_{ki}}{2}a_{ji}a_{ki}\left( a_{jk}+a_{kj}\right) +\\
			&\sum_{j}\sum_{k\neq j}\frac{w_{ij}+w_{ki}}{2}a_{ij}a_{ki}\left( a_{jk}+a_{kj}\right) +
			\sum_{j}\sum_{k\neq j}\frac{w_{ji}+w_{ik}}{2}a_{ij}a_{ki}\left( a_{jk}+a_{kj}\right)&=\\
			&\sum_{j}\sum_{k\neq j}2\frac{w_{ij}+w_{ik}}{2}+\sum_{j}\sum_{k\neq j}2\frac{w_{ji}+w_{ki}}{2}+  \sum_{j}\sum_{k\neq j}2\frac{w_{ij}+w_{ki}}{2}+\sum_{j}\sum_{k\neq j}2\frac{w_{ji}+w_{ik}}{2}&=\\
			&\sum_{j}\sum_{k\neq j}\left( w_{ij}+w_{ik}\right) +\sum_{j}\sum_{k\neq j}\left(w_{ji}+w_{ki}\right)
			+\sum_{j}\sum_{k\neq j}\left( w_{ij}+w_{ki}\right)+\sum_{j}\sum_{k\neq j}\left( w_{ji}+w_{ik}\right)
		\end{split}
	\end{eqnarray*}
	
	\noindent Rearranging the previous expression we obtain:
	\begin{eqnarray*}
		\begin{split}
			\frac{1}{2}[(\textbf{W}+\textbf{W}^T)(\textbf{A}+\textbf{A}^T)^2]_{ii}& =
			2\left( n-2\right)s_{i}^{tot}=\left( 2\left( n-1\right) -2\right) s_{i}^{tot}=\left(d_{i}^{tot}-2\right) s_{i}^{tot}
		\end{split}
	\end{eqnarray*}
	being, for a complete graph, $d_i^{tot}=2\left( n-1\right)$.
	
	\noindent The denominator is:
	\begin{equation*}
		s_{i}^{tot}\left( d_{i}^{tot}-1\right) -2s_{i}^{\leftrightarrow }
	\end{equation*}
	
	Observing that $s_{i}^{\leftrightarrow }=\frac{s_{i}^{tot}}{2}$ for a complete network, then the clustering coefficient is equal to:
	\begin{equation*}
		C^{*}_i=\frac{\left(d_{i}^{tot}-2\right) s_{i}^{tot}}{s_{i}^{tot}\left( d_{i}^{tot}-1\right) -2s_{i}^{\leftrightarrow }}=1
	\end{equation*}
	
	\section{Relation between total coefficient and its specific components}
	
	We report here the computation needed to obtain formula (\ref{Clust_WeightAv}). First, expanding the numerator of (\ref{our_clust}) and rearranging the addends, we obtain:
	
	\begin{equation*}
		\begin{split}
			\left( W+W^{T}\right) \left( A+A^{T}\right) ^{2}=\left( W+W^{T}\right)\left[A^{2}+AA^{T}+A^{T}A+\left( A^{T}\right) ^{2}\right]=\\
			WA^{2}+W^{T}A^{2}+WAA^{T}+W^{T}AA^{T}+WA^{T}A+W^{T}A^{T}A+W\left(A^{T}\right) ^{2}+W^{T}\left( A^{T}\right) ^{2}=\\
			\left[W^{T}A^{2}+W^{T}A^{T}A \right]+\left[WAA^{T}+W\left(A^{T}\right)^{2} \right]+\left[W^{T}AA^{T}+WA^{T}A\right]+\left[WA^{2}+W^{T}\left( A^{T}\right) ^{2}\right]\\
		\end{split}
	\end{equation*}
	
	\noindent The numerator is divided in four addends, each of them equal the numerator of a specific component (i.e. the numerator of formulas ((\ref{in clust}), (\ref{out clust}), (\ref{midd clust}) and (\ref{cycle})). In other words, what emerges by the previous computations is that the total number of actual triangles can be expressed as the sum of all actual triangles of different patterns.
	
	Concerning the denominator, we observe that, by (\ref{tot_deg}) and (\ref{tot_str}):
	
	\begin{equation*}
		\begin{split}
			s_{i}^{tot}(d_{i}^{tot}-1)-2s_{i}^{\leftrightarrow}=\\
			(s_{i}^{in}+s_{i}^{out})(d_{i}^{in}+d_{i}^{out}-1)-2s_{i}^{\leftrightarrow}=\\
			s_{i}^{in}d_{i}^{in}+s_{i}^{out}d_{i}^{in}+s_{i}^{in}d_{i}^{out}+s_{i}^{out}d_{i}^{out}-s_{i}^{in}-s_{i}^{out}-2s_{i}^{\leftrightarrow }=\\
			s_{i}^{in}(d_{i}^{in}-1)+s_{i}^{out}(d_{i}^{out}-1)+s_{i}^{out}d_{i}^{in}+s_{i}^{in}d_{i}^{out}-2s_{i}^{\leftrightarrow }=\\
			s_{i}^{in}(d_{i}^{in}-1)+s_{i}^{out}(d_{i}^{out}-1)+\frac{1}{2}( s_{i}^{out}d_{i}^{in}+s_{i}^{in}d_{i}^{out})-s_{i}^{\leftrightarrow }+\frac{1}{2}( s_{i}^{out}d_{i}^{in}+s_{i}^{in}d_{i}^{out})-s_{i}^{\leftrightarrow }
		\end{split}
	\end{equation*}
	
	Also the denominator is divided in four addends, each of them equal the denominator of a specific component. Hence, the number of possible triangles can be expressed as the sum of all possible triangles of different patterns.
	
	The previous computations entail:
	
	\begin{equation*}
		\begin{split}
			C^{*}_i(\textbf{W})=\frac{\frac{1}{2}\left[(\textbf{W}^{T}(\textbf{A}+\textbf{A}^T)\textbf{A})_{ii}+
				(\textbf{W}(\textbf{A}+\textbf{A}^T)\textbf{A}^{T})_{ii}+(\textbf{W}^{T}\textbf{AA}^{T}+\textbf{W}\textbf{A}^{T}\textbf{A})_{ii}+(\textbf{W}\textbf{A}^{2}+\textbf{W}^{T}\textbf{A}^{T})^{2})_{ii}\right]}{s_{i}^{tot}(d_{i}^{tot}-1)-2s_{i}^{\leftrightarrow}}=\\
			\frac{C^{*,in}_i\left(s_{i}^{in}(d_{i}^{in}-1)\right)+C^{*,out}_i\left(s_{i}^{out}( d_{i}^{out}-1)\right)+
				C^{*,midd}_i\left(\frac{1}{2}\left(s_{i}^{in}d_{i}^{out}+s_{i}^{out}d_{i}^{in}\right)-s_{i}^{\leftrightarrow }\right)+C^{*,cycle}_i\left(\frac{1}{2}\left(s_{i}^{in}d_{i}^{out}+s_{i}^{out}d_{i}^{in}\right)-s_{i}^{\leftrightarrow }\right)}{s_{i}^{tot}(d_{i}^{tot}-1)-2s_{i}^{\leftrightarrow}}
		\end{split}
	\end{equation*}
	that is the formula (\ref{Clust_WeightAv}).
	
	\section*{References}
	
	\bibliography{Myref}

\end{document}